\documentclass[conference]{IEEEtran}
\usepackage[caption=false,font=normalsize,labelfont=sf,textfont=sf]{subfig} 
\IEEEoverridecommandlockouts
\IEEEpubid{\makebox[2\columnwidth][l]{979-8-3503-2277-4/23/\$31.00 ©2023 IEEE CoG\hfill}}
\usepackage{cite}
\usepackage{amsmath,amssymb,amsfonts}
\usepackage{algorithmic}
\usepackage{graphicx}
\usepackage{textcomp}
\usepackage{stfloats}
\usepackage{url} 
\usepackage{xcolor}
\def\BibTeX{{\rm B\kern-.05em{\sc i\kern-.025em b}\kern-.08em
    T\kern-.1667em\lower.7ex\hbox{E}\kern-.125emX}}
\begin{document}

\title{The History of Quantum Games\\
}

\author{
\IEEEauthorblockN{ 1\textsuperscript{st} Laura Piispanen}
\IEEEauthorblockA{\textit{School of Science} \\
\textit{Aalto University}\\
Espoo, Finland \\
0000-0002-5547-9069}
\and
\IEEEauthorblockN{2\textsuperscript{nd} Edward Morrell}
\IEEEauthorblockA{\textit{School of Arts, Design and Architecture} \\
\textit{Aalto University}\\
Espoo, Finland \\
0000-0001-6417-4745}
\and
\IEEEauthorblockN{3\textsuperscript{rd} Solip Park}
\IEEEauthorblockA{\textit{School of Arts, Design and Architecture} \\
\textit{Aalto University}\\
Espoo, Finland \\
0000-0001-5581-435X}
\and
\IEEEauthorblockN{ \makebox[.5\linewidth]{4\textsuperscript{th} Marcel Pfaffhauser}}
\IEEEauthorblockA{\textit{Zurich Research Laboratory} \\
\textit{IBM Research}\\
Zurich, Switzerland}
\and
\IEEEauthorblockN{ 5\textsuperscript{th} Annakaisa Kultima}
\IEEEauthorblockA{\textit{School of Arts, Design and Architecture} \\
\textit{Aalto University}\\
Espoo, Finland \\
0000-0001-5856-5643}
}
\maketitle

\begin{abstract}
In this paper, we explore the historical development of playable quantum physics related games (\textit{\textbf{quantum games}}). For the purpose of this examination, we have collected over 260 quantum games ranging from commercial games, applied and serious games, and games that have been developed at quantum themed game jams and educational courses. We provide an overview of the journey of quantum games across three dimensions: \textit{the perceivable dimension of quantum physics, the dimension of scientific purposes, and the dimension of quantum technologies}. We then further reflect on the definition of quantum games and its implications. While motivations behind developing quantum games have typically been educational or academic, themes related to quantum physics have begun to be more broadly utilised across a range of commercial games. In addition, as the availability of quantum computer hardware has grown, entirely new variants of quantum games have emerged to take advantage of these machines' inherent capabilities, \textit{quantum computer games}.
\end{abstract}
\section{Introduction}
\textit{Quantum games} refer to the video games, analogue games, and all other variant of games that \textit{reference the theory of quantum physics, quantum technologies, or quantum computing through perceivable means, connect to quantum physics through a scientific purpose or use quantum technologies.} Whilst the earliest quantum games date back to the early 1980's \cite{piispanen2022}, the number of games relating to quantum physics and quantum technology has greatly increased in the last few years, as interest has grown around their academic and educational use. It is therefore essential that the breadth of quantum physics related games is established, to provide the cultural background, technical frameworks and artistic inspiration relevant for the future development. It should be acknowledged that a significant number of quantum physics related games are not strictly educational, academic, or else created with serious or applied purpose, nor do they often have use for any such application. In providing a history of quantum games, inclusive of games for entertainment, we present this work in the context of the expanding \textit{ludosphere} \cite{stenros2018}, recognising the value of mapping specific game histories despite the ever increasing volume of games being made. In the absence of any prior systematic account of quantum game history, this paper provides a necessary foundation for the future work and play of quantum games.

For the purpose of understanding the history of quantum games, we have collected over 260 games with varying degrees of reference to quantum physics \cite{quantumgames}. In this paper, we have utilised a range of these samples to overview the beginning and development of quantum games. These games are presented through aspects referenced as the \textit{dimensions of quantum games} (See Table \ref{table:definition}), which are sought either directly through the gameplay or provided externally through developer interaction, or via additional paratexts such as rule books \cite{consalvo2007}.

\begin{table}[!ht]
    \centering
    \caption{The dimensions of quantum games as described by Piispanen et al. (2023) \cite{piispanen2022}.}
    \label{table:definition}
    \begin{tabular}{|l|l|}
    \hline
        \textbf{DIMENSIONS OF QUANTUM GAMES} \\ \hline
        \textbf{Perceivable dimension of quantum physics:} \\ 
        The reference to quantum physics in the game is perceivable by \\
        interacting with the game or with its peripheral material \\
        \hline
        \textbf{Dimension of quantum technologies:} \\
        The game incorporates usage of quantum software or actual quantum \\ 
        devices either during the gameplay itself, or during the development \\ of the game \\ \hline
        \textbf{Dimension of scientific purposes:} \\
        The game is intended to be an educational game, a citizen science \\
        game, uses a tool designed for such games or otherwise has a purpose \\
        towards a scientific use \\ \hline
    \end{tabular}
\end{table}

Prior academic work detailing the concept of quantum games has typically focused on exploring only those games with a serious purpose \cite{goff2006, gordon2010, cantwell2019, jensen2021, seskir2022}, which can give the impression that quantum games are always intended as educational or scientific \cite{marfisi2010,marklund2013, seriousgames}. Here, we expand upon this notion by looking at the entire ludosphere through the many different motivations that have inspired (quantum) game developers. Beyond the aforementioned serious use cases, common motivations we found include testing the abilities of quantum computers and utilising quantum physics to provide thematic or mechanical flavour for games created in both commercial contexts and as part of jams or hackathons.

\textit{The List of Quantum Games} at \cite{quantumgames} has been collected and regularly updated since 2017, utilising data collected through the Quantum Game Jam (QGJ) events \cite{kultima2021qgj}, academic papers, public preprints, academic conferences, and public outreach projects. The list was collated with the help of the community of quantum game developers and quantum physics enthusiasts, and other interested parties. The document is publicly accessible to enable in the assistance of future quantum games research, and the authors welcome to any inquiries or suggestions. 

\section{Context: From the first computers to quantum technology }

\begin{figure}[ht]
\center 
\subfloat[]{\includegraphics[width=0.6\linewidth]{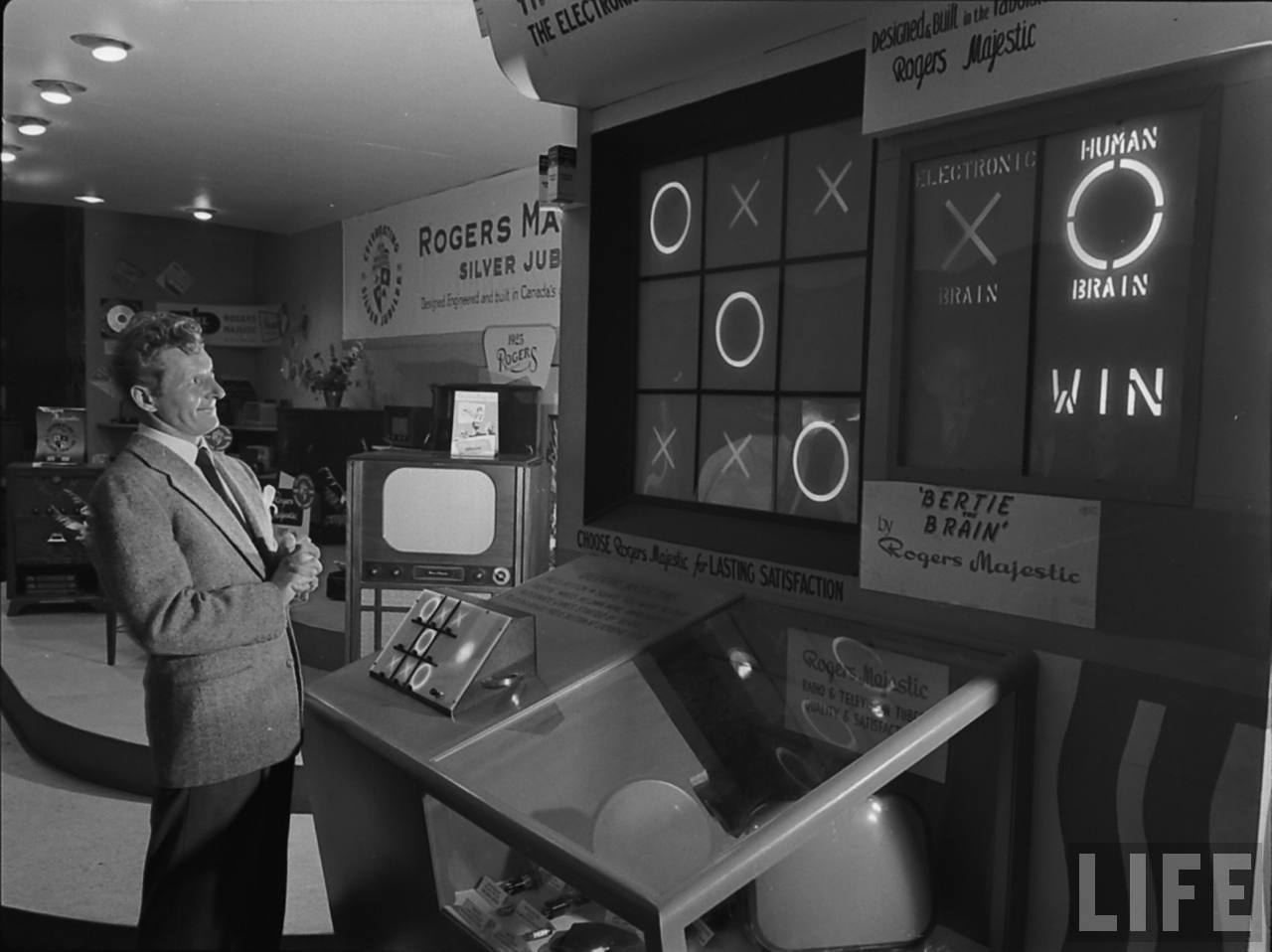}}
\subfloat[]{\includegraphics[width=0.35\linewidth]{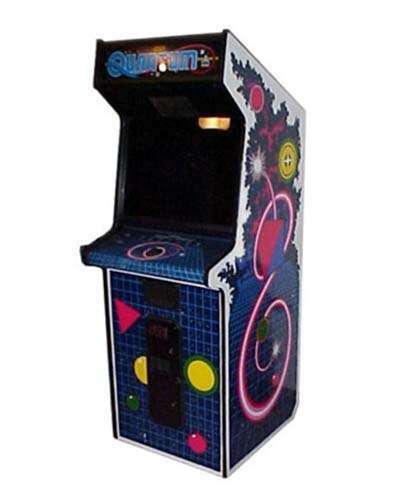}}
\caption{a) Life magazine photo of comedian Danny Kaye standing in front of Bertie the Brain at the Canadian National Exhibition in 1950. (Bernard Hoffman 1950) b) A picture of the Atari's \textit{Quantum} arcade from 1983 (Joystick).}
\label{fig:bertiespacewar}
\end{figure}

Computer games have existed for nearly as long as there have been computers, a history dating back to the 1950's with electronic games such as \textit{Bertie the Brain} (See Figure \ref{fig:bertiespacewar}a ), designed to demonstrate the use of light bulbs and vacuum tubes by playing \textit{Tic-tac-toe}. Programming principles and algorithms were then a new and confusing subject, so well-known, traditional games were often used as an educational tool to illustrate the possibilities of computing. As such, games developed for early computers were often about showcasing new technologies, educating people about them, or researching the capabilities of the technology itself. In Section \ref{sec:qcg} we illustrate the first games on quantum computers, namely \textit{quantum computer games}, sprung out of similar motivations.

Quantum physics is a theory that describes the behavior of particles at the subatomic scale, encompassing phenomena such as quantum superposition, quantum measurement, and quantum entanglement -- phenomena not observed in our everyday lives \cite{busch1995, zeilinger1999}. Quantum technologies, which rely on these quantum phenomena, have emerged as a new market with significant potential \cite{dowling2003, deutsch2020}. A prominent example of quantum technologies is the development of quantum computers, which exploit quantum physical phenomena to manipulate and store information \cite{benioff1980, deutsch1985, divincenzo1998, divincenzo2000, nielsen2002, ladd2010}. In contrast to classical computers that use bits, quantum computers employ quantum bits, or \textit{qubits}\footnote{The spelling "qbits" is also used.}. Qubits can exist in a state of quantum superposition, simultaneously representing both 1 and 0, with a quantum measurement collapsing the state to a definite value. Additionally, qubits can be entangled, strongly correlated even over long distances, so that the measurement of one qubit reveals information about the others. These unique properties of qubits enable fundamentally different and powerful methods of information processing.

There exist complex problems that cannot be effectively simulated or solved using even the most powerful classical computers. These include breaking current cryptography, modeling intricate molecular structures for medicine, efficient database searches, and various optimization problems \cite{grover1996, robert2021, shor1994, horowitz2019, outeiral2021, quantumalgorithmzoo}. While supercomputers can partially address such problems and simulate small-scale quantum computers, the computational resources required grow exponentially with the number of qubits \cite{zhou2020}. Therefore, the development of scalable and reliable quantum computers is crucial. It is worth noting that quantum technologies are not expected to completely replace classical computers, and that they have already contributed to numerous inventions such as lasers, transistors, superconducting materials, and the miniaturized computers found in mobile phones and smartwatches today.

\section{Quantum Games}
In this section we explore the history of quantum games, utilising a definition of quantum games \cite{piispanen2022}, formulated via an analysis of over 250 games \cite{quantumgames}. The history is presented according to the reasoning that a quantum game can be defined based on three dimensions, namely the \textit{perceivable dimension of quantum physics}, the \textit{dimension of scientific purposes}, and the \textit{dimension of quantum technologies} \cite{piispanen2022}.

\subsection{Early days of Quantum Games}
The earliest example of a quantum game found in our data is Atari's vector based arcade game \textit{Quantum}, released in 1982 (See Figure \ref{fig:bertiespacewar}b ). The game is described by the creators as being set in the 'sub-atomic world of quantum physics', with the player using an optical trackball to capture various particle types without clashing to them. This was categorised as a quantum game owing the reference to phenomena at the atomic scale.

In the following three decades, stretching from the golden age of arcade video games in the 1980s towards the networked and accessible game development environments of today, only a small number of games fitting the definition of a quantum game have been recorded \cite{quantumgames}. Any omissions from this history would be owing to the colossal volume of games being regularly produced across multiple languages, a limitation that we recognise and contextualise by positioning this history within the larger \textit{ludosphere} \cite{stenros2018}.  

The Australian science fiction writer and amateur mathematician Greg Egan published \textit{Quantum Soccer} on his personal website in 1999, presenting the game alongside a work of fiction and the mathematical details determining the rules. The game has the player trying to shape the wave function of a quantum-mechanical 'ball' to increase the possibility of it being inside a goal. In presenting a simulated visualisation of a wave function through game mechanics, \textit{Quantum Soccer} fits the definition of a quantum game through a perceivable dimension.

The first mobile game labelled as a quantum game was released in 2009 on the Apple App Store, \textit{Universe Splitter}. The mechanics of the game vary little from a coin toss, with the game telling you which choice to make when provided with any two actions by the player. However, the game claims to determine that choice by connecting to a quantum device in Geneva to release single photons into a partially-silvered mirror which, according to the many-worlds interpretation of quantum mechanics, would result in them existing in separate universes. \textit{The dimension of quantum technologies} is possible, though not verified in this game, but through the reference to the many worlds interpretation, it has \textit{a perceivable dimension of quantum physics} in it.

\textit{Quantum Race} was presented in 2011 at the 'Festival della Scienza di Genova', a giant live-action board game that was specifically designed to explain complex science concepts to school children \cite{chiarello2015}. The principles of quantum mechanics, including wave functions, delocalisation, collapse and the tunnelling effect are all present in the game's rules, with the players controlling cars that split apart and collapse as they race each other. Similarly engaging a younger audience with an accessible introduction to quantum principles is the mod \textit{qCraft}, released in 2013 for the game \textit{Minecraft}. Rather than simulating quantum physics, the mod presents 'analogies' for the player that can demonstrate how quantum behaviours vary from everyday expectations. Both of these examples have a strong \textit{dimension of scientific purposes} through their educational nature alongside a clearly perceivable layer of quantum physics. 

\begin{figure}
\subfloat[]{\includegraphics[width =0.5\linewidth]{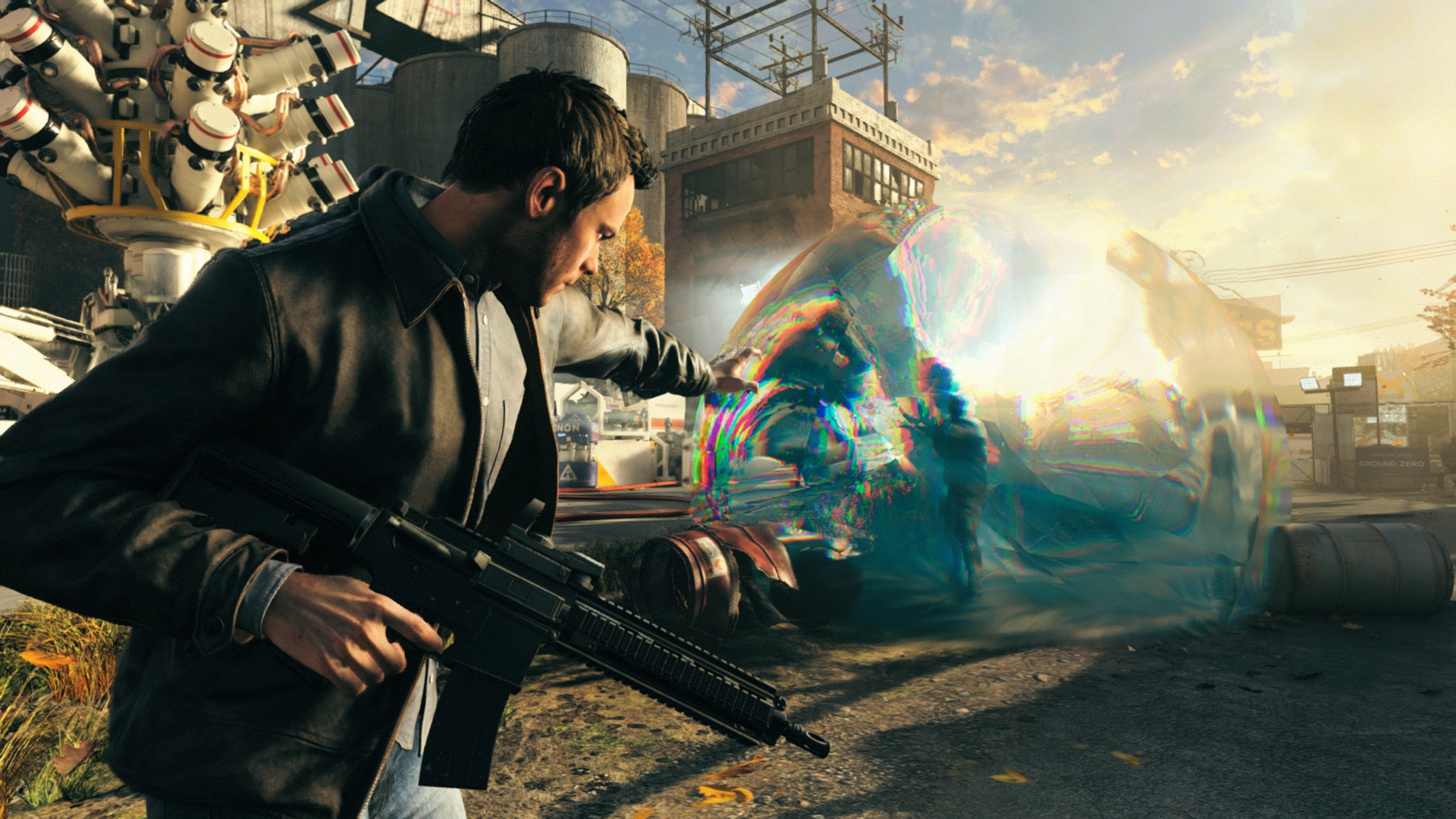}\,
\includegraphics[width = 0.5\linewidth]{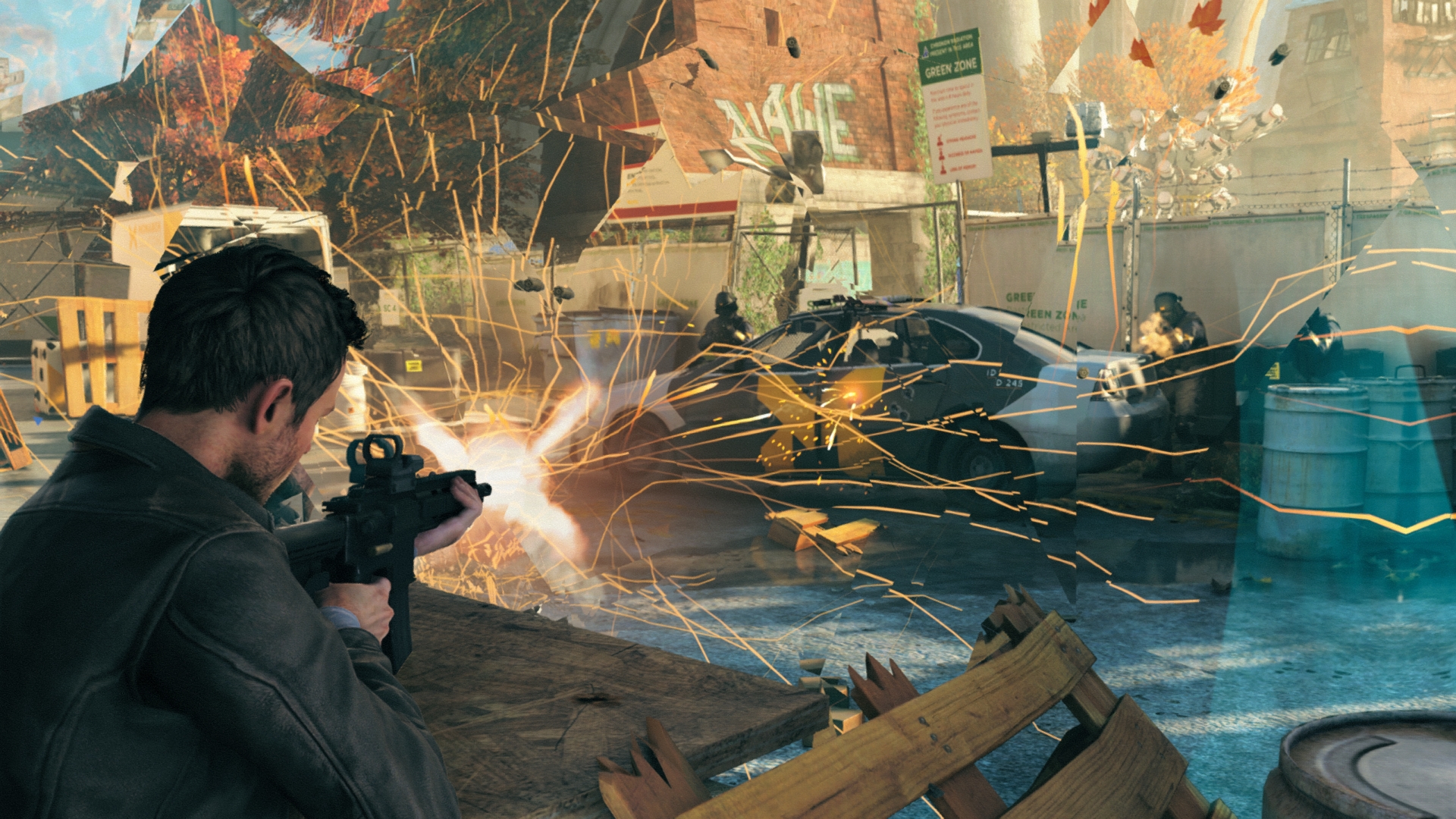}}\,
\subfloat[]{
\includegraphics[width = 0.5\linewidth]{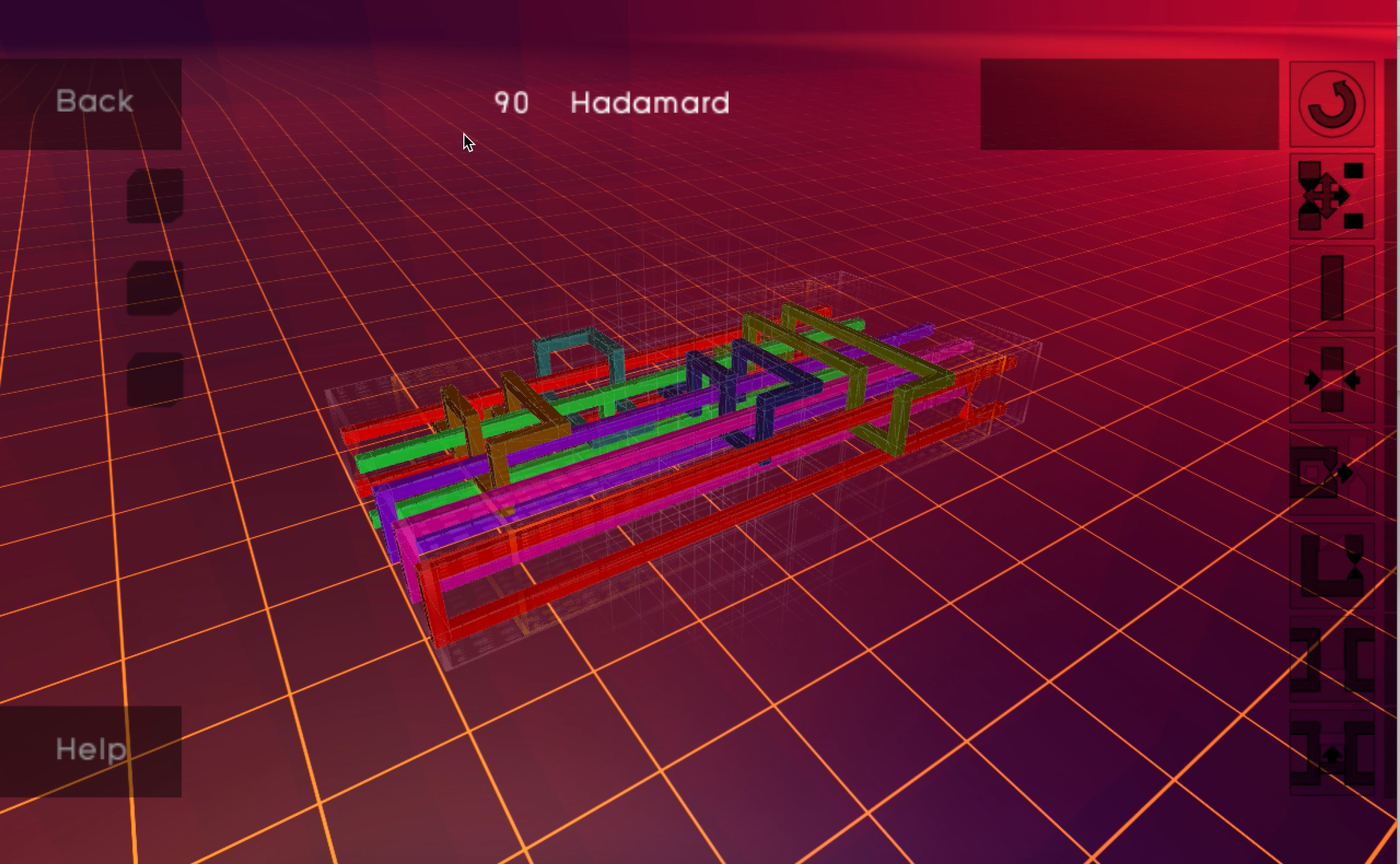}\,
\includegraphics[width = 0.5\linewidth]{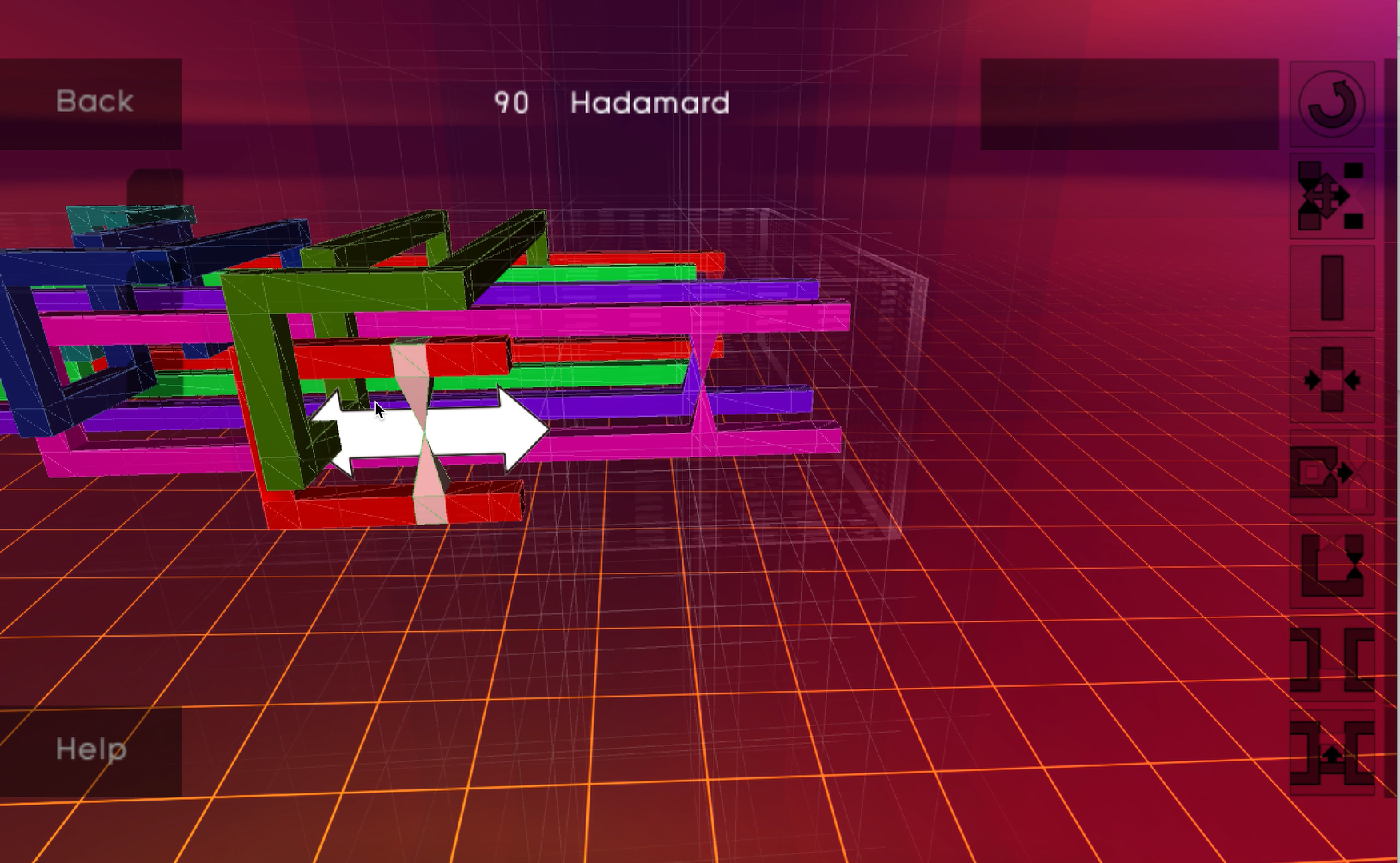}}\\
\caption{Screenshots taken from games (a) \textit{Quantum Break} (Remedy Entertainment, Press kit), (b) 
and from \textit{meQuanics}. \textit{Quantum Break} is a science fiction action-adventure third-person shooter video game, where the protagonist is able to control the flow of time. In the citizen science game \textit{meQuanics} the player solves puzzles consisting of complex knotlike structures.}
\label{screenshots00}
\end{figure}

It has to be mentioned that not every game that references the word 'Quantum' can be considered a quantum game \cite{piispanen2022}. For example, in 1989 Cosmicsoft published the game \textit{Quantum} which had no discernable relevance to quantum themes, instead being a rebranding exercise for a game previously known in 1988 as \textit{Game Over II} and \textit{Phantis}. Far more likely is that the name was inspired by the hit television series \textit{Quantum Leap}, that had launched earlier that same year. Across all forms of entertainment media there are examples of the term 'Quantum' or more particular examples of quantum physics being used to boost science-fiction themes, with the science, more often that not, quickly replaced with pure fantasy. From the examination of Piispanen et. al \cite{piispanen2022}, to classify as a quantum game, there would need to be a legitimacy to the quantum physics connection, either through the game play or peripheral material. During the development of 2016's \textit{Quantum Break}, a quantum physicist was consulted to connect the actions and visuals to quantum theory (See Figure \ref{screenshots00}a ) \cite{qbreak}, lending legitimacy to the \textit{perceivable dimension of quantum physics} references, that led Piispanen et al. (2023) to classify \textit{Quantum Break} as a quantum game \cite{piispanen2022}.

\subsection{Games for Quantum physics}
Digital games have the potential to offer far more then just a superficial incorporation of quantum physics theory. The principles behind the theory has traditionally been taught in universities through books, lectures and exercise sessions using mathematical formalism. As quantum phenomena offer no real-life reference, studying these abstract concepts can be challenging and may lead even to misconceptions \cite{styer1996, singh2001, singh2007, zhu2012}. For a long time, digital computers have been used for creating interactive tools for simulating and visualising quantum physical phenomena \cite{brandt1989, mckagan2008, kohnle2012, irmak2023} and for educational purposes to strengthen the learning and mental model building processes \cite{passante2019, cataloglu2002, keebaugh2019, muller2002, anupam2019}. 

With the use of simulations and designing the rules of a game in a diligent manner, it is possible to incorporate quantum phenomena into an existing game framework in a refined way. This allows the player to interact with simulated quantum phenomena rather than fiction and misconceptions \cite{goff2006,gordon2010,cantwell2019}. One of the earliest examples of such games was proposed by Allan Goff in 2002, with his idea of \textit{Quantum Tic-Tac-Toe} \cite{goff2006}, inspiring the creation of several variations of the classic game, such as \textit{Quantum TiqTaqToe} in 2019 by Evert van Nieuwenburg. 

Games have been seen as an approachable medium to utilise simulations, and in particular, they demonstrate a lot of promise in introducing new concepts and teaching the basics of quantum mechanics in a playful, immersive manner. This has led to the development of educational games and learning platforms integrating games with learning materials \cite{cantwell2019, helloquantum, quantime, sbasedgames, scienceathome, qplaylearn}. It is evident that quantum technologies offer advantages over existing technologies \cite{preskill2018}, and thus large investments have been made worldwide into developing technologies dependent on quantum phenomena. This in turn has increased demand for a trained workforce into the growing quantum technology industry field \cite{fox2020}, demands that will continue to grow in the future. As such, the demand for quantum games as one of the new teaching methods to teach quantum physics has vastly grown in recent years, particularly in European and US education \cite{seskir2022,q2work}.

Educational games about quantum physics have a \textit{dimension of scientific purpose} \cite{piispanen2022} and are therefore categorized as one of the key variants of quantum games. This dimension partly coincides with the earlier definition of "\textit{quantum computer games}” by Gordon \& Gordon from 2010, that provided an easy distinction between games on computers and \textit{Quantum Game Theory}, a theoretic extension to quantum physics concerning the mathematical study of optimising rational agents \cite{gordon2010,wiesner1983,deutsch1992,meyer1999,eisert1999}. In this paper, we have chosen to instead assign the descriptor “quantum computer game” specifically to games \textit{on} quantum computers (by referencing \cite{piispanen2022}, as discussed in Section \ref{sec:qcg}). Additionally, the previous definition by Gordon \& Gordon covered only educational games, which served the purpose of introducing their game, the \textit{Quantum Minesweeper}. We believe that labeling educational games under the sub-variant of \textit{dimension for scientific purposes} will enable richer discourse to diversify the modalities of the definition of quantum games.

Citizen science games related to quantum physics research are a sub-variant of the \textit{dimension of scientific purposes} \cite{piispanen2022}. Games have provided a powerful tool for citizen science in, for example, the study of protein structures and galaxy classification \cite{cooper2010, raddick2009}. The first quantum physics-related citizen science game called \textit{Quantum Moves} was developed in Denmark between 2011 and 2012 by the research group "Science at Home", which lead to the development of \textit{Quantum Moves 2} published in 2018 \cite{jensen2021, sherson2022} (See Figure \ref{fig:catboxscissors}b ). These games are centred around the problem of optimising a quantum state-transfer processes within the framework of quantum optimal control \cite{dowling2003, koch2016}. Other research topics in quantum physical sciences that have motivated the development of citizen science games include testing the fundamental theory of quantum information, quantum algorithm development, and quantum error correction. \textit{Bell's theorem} by Kaitos Games, Spain 2016, collected data from more than 100,000 players in order to create human-sourced randomness for experiments like a version of Bell’s test, a theorem indicating the non-local characteristics of quantum entanglement \cite{bellgame2018}. The citizen science game \textit{meQuanics} by h-bar was developed in 2016 in Australia for quantum algorithm optimising, through the action of distanglening interconnected loops (See Figure \ref{screenshots00}c )  \cite{devitt2016}. There, the player aims to manipulating complex knot-like structures restrained by the rules given in the game. \textit{Decodoku} by James Wootton from 2018 was a game that provided little puzzles, solving of which lead to solutions for quantum error correction \cite{wootton2017}.  

\subsection{Games on Quantum Computers}
\label{sec:qcg}

\begin{figure}[ht]
\center 
\subfloat[]{\includegraphics[width=0.5\linewidth]{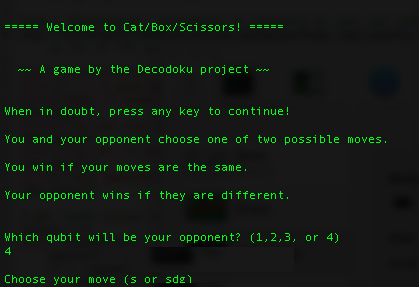}}\, 
\subfloat[]{\includegraphics[width=0.65\linewidth]{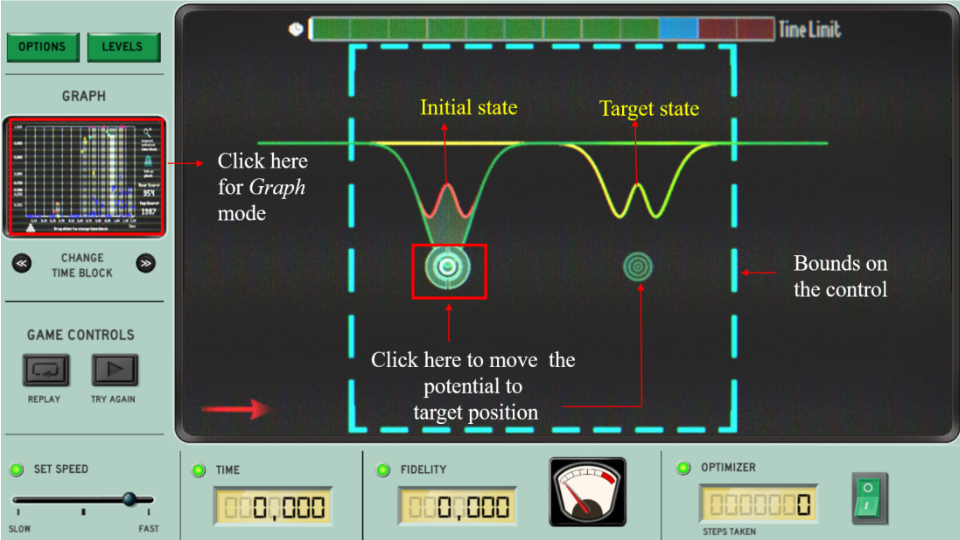}}
\caption{ Screenshots of a) the quantum computer game \textit{Cat/Box/Scissors} by James Wootton and b) \textit{Quantum Moves 2} by Science at Home.}
\label{fig:catboxscissors}
\end{figure}

Some of the early devices that could be considered as quantum computers have been publicly accessible for researchers and quantum enthusiasts via the IBM Q-Experience\footnote{https://quantum-computing.ibm.com/} since 2016 through a graphical user interface. The first quantum game running on a quantum computer, a \textit{quantum computer game} \cite{piispanen2022}, was released the next year in 2017 by James Wootton. Titled \textit{Cat/Box/Scissors}, the game was developed using the Python computer language as a demonstration of the first available quantum devices, albeit through a simple command line interface (See Figure \ref{fig:catboxscissors}a ) 
\cite{woottonhistory}. The game allows you to play against an opponent, whose moves are affected by quantum-sourced randomness that leads to noise on the opponent’s moves, offering the possibility of a winning tactic over the quantum computer. The \textit{Cat/Box/Scissors} was inspired by the traditional game of \textit{Rock-Paper-Scissors}, and led the creator to developing the first multiplayer game for quantum computers in 2017. The game \textit{Quantum Battleship} is another command line game that is inspired by the game of \textit{Battleships}. It takes advantage of entanglement between two qubits and incorporates the phenomena of quantum coherence to the game play. 

Soon after the release of \textit{Cat/Box/Scissors}, IBM released an open-source software development kit called \textit{Qiskit}, which allowed developing quantum programs for quantum devices on a local computer through simulations of quantum computers. This, combined with the cloud service access to these early quantum devices, was followed by papers about adapting quantum game theoretic procedures on the devices \cite{becker2019, sagole2019, chowhan2020}. This sometimes results in a playable game in the form of conventional games --- not just video games but board games, with properties added from the quantum physical theory \cite{du2002,sagole2019}. However, it should be noted that these implementations have not been specifically developed for gaming purposes but were instead specifically developed to experiment on quantum game theory and showcase what advantages implementing quantum strategies provides for game strategies. 

Many of these implementations are openly accessible and playable through the Jupyter platform, or with pen and paper when combined with the IBM Q-Experience interface. These games are quantum games in both the context defined here through the \textit{perceivable dimension of quantum physics} (game rules) and through the \textit{dimension of quantum technologies} (quantum computers) \cite{piispanen2022}. Those that were specifically introduced as pedagogic tools or as a “tangible intro to quantum phenomena” \cite{chowhan2020} can also be considered as having a \textit{dimension of scientific purposes} (education). 

The cloud-based services offering access to quantum computers rely on holding a queue of tasks from different users, which can make the call of a quantum computer a lengthy wait. This is a challenge for quantum computer game development, but it can be bypassed by sourcing (certain) data beforehand. The quantum computer game, \textit{Quantum Solitaire} by James Wootton from 2017 takes advantage of such a call procedure. The game was developed using Unity and is playable online, but it can be said that it is also running on a quantum computer, thus displaying the \textit{dimension of quantum technologies}. The connection to quantum computers remains a novel resource, so it is understandable that it is underlined through text prompts in the game, incorporating also a \textit{perceivable dimension of quantum physics}. Succeeding in \textit{Quantum Solitaire} does not require any previous knowledge of quantum physics, but the game claims to have the possibility of providing “an intuitive feel of what quantum entanglement can do” \cite{wootton2017m}.\\

Several companies have also developed quantum computer games, again, mostly for educational purposes. In 2018 IBM Research released the board game \textit{Entanglion} and the mobile game \textit{Hello Quantum} to introduce basic quantum computing concepts \cite{entanglion, helloquantum, wootton2020teach}. Meanwhile, \textit{C.L.A.Y.} from Mitale Games claims to be the first commercial quantum computer game \cite{piispanen2022}. While it contains no \textit{perceivable dimension of quantum physics} except through its marketing material, and also contains no \textit{dimension of scientific purposes}, it runs subroutines on a quantum computer simulator and on quantum hardware \cite{clay2019}. \textit{The Qubit Game} by Google Quantum AI provides educational content about the workings of quantum computers.

\subsection{The Rise of Quantum Game Development}
\begin{figure}
\subfloat[]{\includegraphics[width = 1.6in]{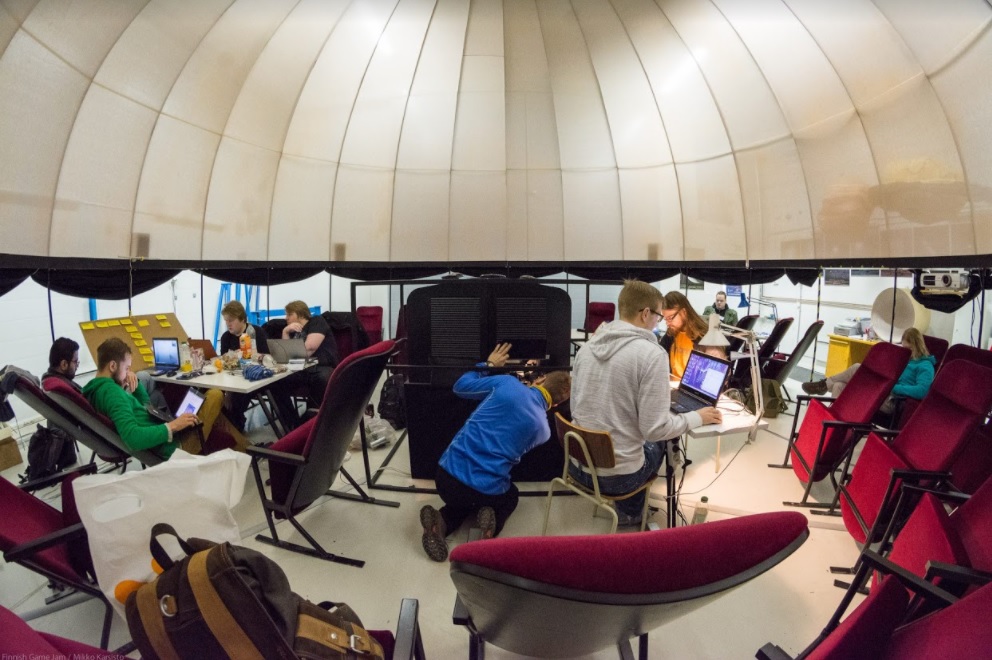}}\,
\subfloat[]{\includegraphics[width = 1.9in]{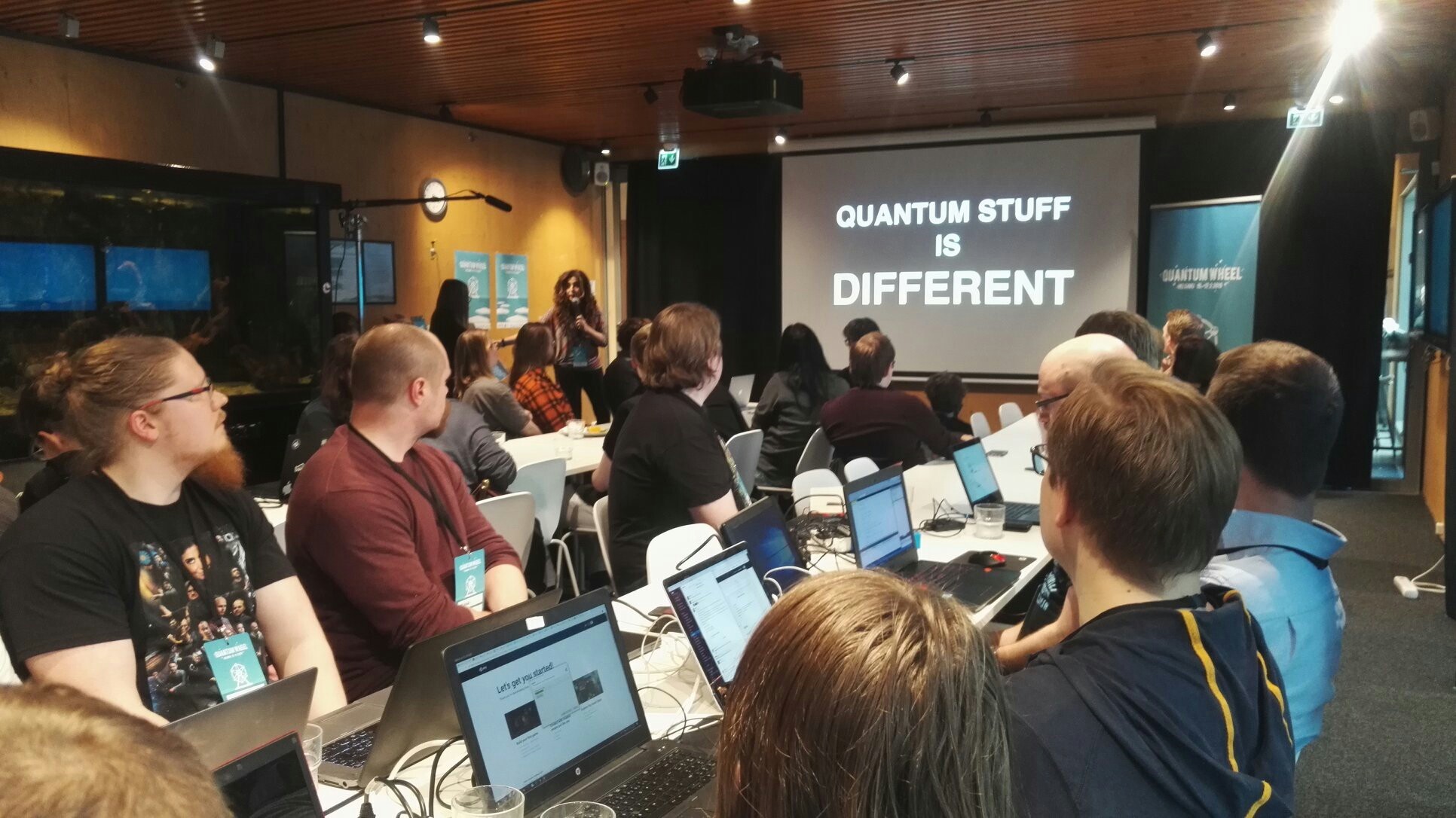}}\\
\caption{a) A picture taken at the first ever QGJ in 2014. The event was situated at an observatory and some games were developed to be played on the dome (\textit{Jaakko Vainio}). b) Picture taken at the QGJ of 2019, \textit{Quantum Wheel}. At the beginning of a QGJ, a presentation is given about the involved technologies and quantum physics (\textit{Laura Piispanen}). }
\label{fig:gamejams}
\end{figure}

We have thus far largely focused on games with serious purposes created by academics, but when considering the \textit{List of quantum games} we notice that approximately half of the games listed were created during game jams and hackathons \cite{quantumgames,kultima2015}. In 2014, an annual series of 6 events called the \textit{Quantum Game Jam} (QGJ) began with the goal of bringing together people both from quantum physics research and professional game development (See Figure \ref{fig:gamejams}) \cite{kultima2021qgj}, and has continued since then as an online event \cite{piispanen2023qgj}.  

Some of the developed games have only had an inspirational, \textit{perceivable dimension of quantum physics} such as the game \textit{What’s in the Box?} from the year 2015 QGJ by unknown creators from Tempe, USA. The game has the player fight against a cardboard box in a street fighter setting. The player is able to look in the box any time during the fight and take their chances that the opponent inside it is beaten enough. The serious purpose for developing quantum games at hackathons also contributes to the number of quantum games with a \textit{scientific dimension of quantum games} \cite{piispanen2022}.

QGJs, in particular, have also resulted in games that have no \textit{perceivable dimension of quantum physics}, but may still for example, use a numerical simulation of a quantum physical system either in the development phase or throughout the run of the game \cite{piispanen2022}. The latter coincides with the \textit{dimension of quantum technologies} and sometimes also with the \textit{dimension of scientific purposes}, for instance when a particular numerical simulation offered at the events is utilised to aid in the development of citizen science game prototypes \cite{kultima2021qgj}.

\begin{figure}
\includegraphics[width = 1\linewidth, height=1.8in]{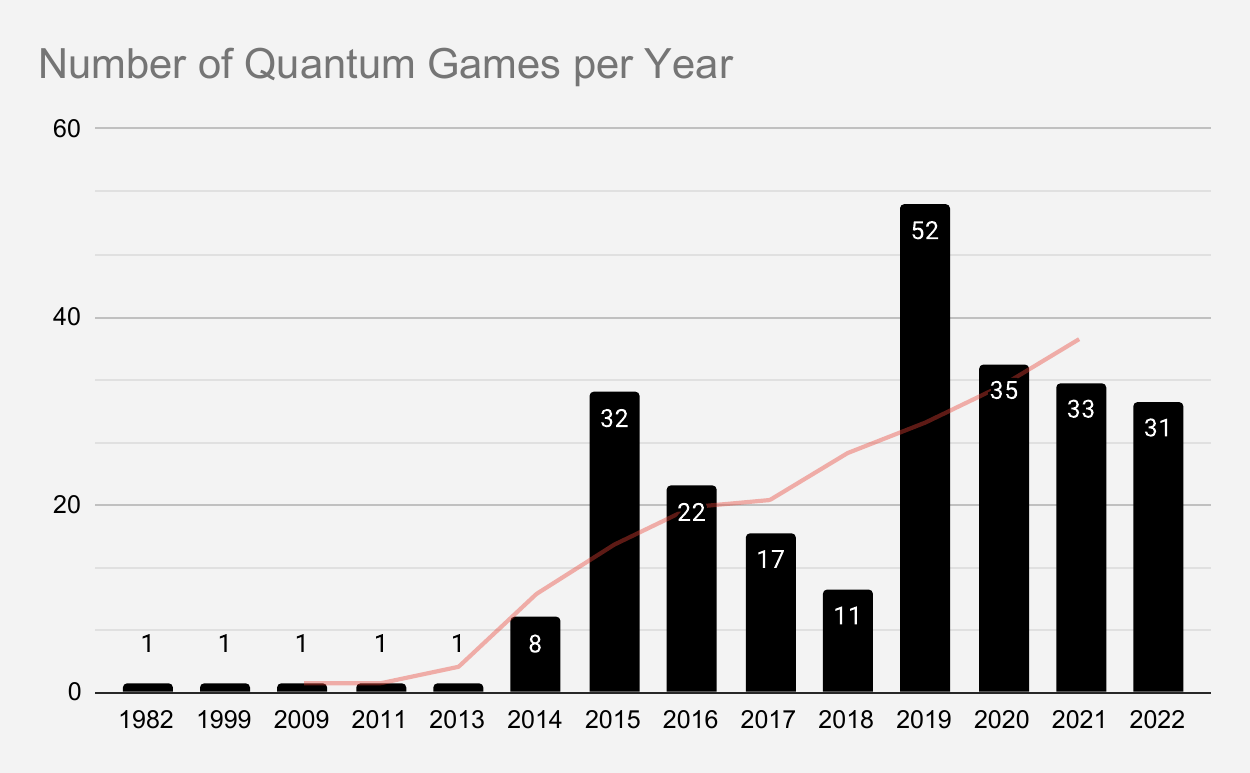}\,
\caption{A table showcasing the number of quantum games included in the \textit{List of Quantum Games} and the moving average of these numbers (a statistic that captures the average change in a data series over time). \cite{quantumgames}.}
\label{fig:gamesperyear}
\end{figure}

Quantum game jam events and hackathons have continued to motivate people from outside academic circles to participate in quantum game development. There is a peak in the number of quantum games developed in 2015 (See Figure \ref{fig:gamesperyear}), which is mostly attributed to a global edition of the QGJ emphasising serious game development. The events attracted many other quantum science research groups around the world, leading to a total of 28 games being developed in the event that year \cite{kultima2021qgj}. The motivation of developing citizen science game prototypes was carried in the themes of QGJs until 2019. 

IBM, Google AI and Microsoft offer tutorials on developing on quantum computers \cite{helloquantum, woottonHello, quantumai}. The goal in offering such materials has been to involve many talented people in the open source development frameworks and assistance with developing games, which in particular was seen as a promising way to introduce the idea of quantum computing \cite{wootton2017m}. In 2019, QGJ and IBM Research organised a special event aimed at experienced game jammers and game developers, the \textit{Quantum Wheel} \cite{kultima2021qgj}, the first jam to encourage the use of quantum computers. IBM Research organised dedicated cloud access to their quantum computers alongside mentoring for the participants, to support development of quantum computer games that could aid in the exploration of the capabilities of quantum computers \cite{wootton2017m, wootton2018}. Half of the presented games from \textit{Quantum Wheel} were running either simulations of IBM’s quantum computers or calculations directly on them, while the other half ran the citizen science motivated simulation tool \cite{kultima2021qgj}. Whilst some of the games were using the provided resources purely as random number generators, two of the games from this jam were later further developed towards the use of citizen science and education, \textit{QCards} and \textit{Hamsterwave} \cite{qplaylearn, piispanen2023projects}.\\

In addition to the QGJ of 2019, events like the IndiQ QgJ, Qiskit Camp 2019, Qiskit Hackathon Madrid, Qiskit Camp Europe, Qiskit Hackathon in Singapore, Quantum Futures Hackathon, Qiskit Camp Asia, CSS Qiskit Hackathon, Qiskit Camp Africa and IBM Q Awards 2019 were all able to attract interest toward combining simulated or quantum sourced quantum computational aspects with games during the same year \cite{indiq}. This can be seen as the second spike in the number of quantum games developed per year (See Figure \ref{fig:gamesperyear}). The series of QGJs and various quantum game hackathons have since 2020 been organised mostly as online events, supporting the global collaboration of quantum games \cite{piispanen2023qgj}.

\section{Discussion}
Quantum games have been created for the purposes of science fiction, education, citizen science and exploring the possibilities of quantum computing and quantum computers. While the history presented here is comprehensive as a foundational work, we acknowledge that the \textit{List of Quantum Games} \cite{quantumgames} we have used as a primary source contains missing data that could add to this picture, lacking comprehensive information on the years of publication, creators, and origins of each of the included games. The authors have encountered difficulties in obtaining this information, with some games taken offline or never publicly accessible from the outset. We still expect that few of the listed games could have been created before the year 2014. The list may also be lacking potentially relevant quantum games that do not reference quantum physics or related phenomena in their title or marketing, or are otherwise more ambiguous in their utilisation of the concept. It is appropriate to point that the definition of quantum games provided by Piispanen et al. \cite{piispanen2022} relies mostly on the games developed between the years 2014 and 2022. This definition is likely to be suited to any quantum games older than this sample range, but we expect future implementations of quantum technology to go beyond what we can categorise today.

From the data we notice that there are five participants who have been credited in at least five quantum games. From these five, one is a company, two are quantum physicists and two are experienced game jammers/developers. With the exception of the listed company, all of these creators have participated in QGJs, but only 13 of their 29 quantum games were created during a QGJ. In addition to these five creators, there are 10 others credited in at least 3 quantum games. Overall, there are over 400 unique names, nicknames and company names credited for the creation of quantum games and just under 400 of them have one quantum game to their credit. 

Many of the quantum games can be labelled in the Puzzle genre. Some of these puzzles require some knowledge about the structure and rules of quantum physics and some aim at teaching this logic to the player, either through a tutorial or in the gameplay. It is important to notice though, that many of the existing quantum games do not have a \textit{dimension of scientific purposes} \cite{piispanen2022}, meaning that they have not been intended for any serious purpose like the education of quantum physics. No profound studies exploring even the majority of existing quantum games from the perspective of their potential for educational or outreach purposes have yet been conducted \cite{piispanen2022, seskir2022}, but the \textit{dimensions of quantum games} could very well offer a starting point for the analysis of these aspects \cite{piispanen2022}. \\

It should also be noted that even the most carefully designed educational game cannot be assumed to work as a stand-alone tool that could teach the multifaceted nature of quantum physics to the player, without extensive user testing \cite{kultima_designing_2020}. The language and foundation of quantum physics is mathematical and complex, requiring a certain level of rigour to understand, which should be separately evaluated. Although games and simulations have proven to be beneficial in teaching some aspects of quantum physics, they still require supervision and monitoring from a quantum expert or a specialist \cite{ahmed2021}.

We have found the \textit{dimensions of quantum games} proposed by Piispanen et al. \cite{piispanen2022} to be valuable in the analysis and evaluation of quantum games. We recommend further investigation and development of these dimensions to explore different strategies for supporting the teaching and outreach of quantum physics and quantum computing. Some quantum games may be helpful for an interested individual to take their first glimpse into the world of quantum physical phenomena or the logic of quantum computing, while other games might provide an interactive way to explore simulations in a classroom setting. Another important layer for educational games is the \textit{perceivable dimension of quantum physics}. Along this dimension it would be essential to make sure that the spoken and written content are justifiable and true, and that any visual queues or dynamics are based on purposeful simulations and references. The third dimension of quantum games, the \textit{dimension of quantum technologies}, becomes an important subject when finding the suitable connections to numerical simulations and when demonstrating aspects of quantum computing. 

We acknowledge that quantum computers provide the connection between playable quantum games and quantum game theoretic quantum strategies. Implementations of games like the \textit{Tic-Tac-Toe, Chess, Go} etc. with expanded rules to incorporate the use of quantum phenomena can lead to quantum games that fit all three dimensions of quantum games \cite{piispanen2022}. Yet, for a game to be immersive and relevant, especially when aiming to educate,  it likely requires more than a command line or Jupyter interface.

\section{Conclusion}
The development of quantum games has received a lot of interest from research groups involved in quantum technology development, quantum computing experts and quantum physics educators, mostly for serious game purposes. Yet, especially since the formation of \textit{Quantum Game Jams} \cite{kultima2021qgj}, artists and game developers with no previous background in quantum physics theory have played a significant role in the expansion of quantum games \cite{piispanen2022}. Not all quantum games have a designed serious or applied purpose, with quantum physics also inspiring entertainment games both commercially as well as through smaller projects. It has been essential to establish the history of quantum games so that their purpose and potential can be better understood when utilised in education or training. Beyond this, attractive quantum games and interactive quantum art will play a vital role in further popularising the appeal of quantum physics to the general public. While the future of quantum games could likely be beyond all our comprehension, the past started here. 

\begin{IEEEkeywords}
Serious Games, Applied Games, Quantum Physics
\end{IEEEkeywords}

\section*{Acknowledgment}
LJP would like to acknowledge their research has been funded by the use of Academy of Finland PROFI funding under the Academy decision number 318937 and by the Vilho, Yrjö and Kalle Väisälä Foundation of the Finnish Academy of Science and Letters. 

\bibliographystyle{ieeetr}
\bibliography{references}  

\end{document}